\documentstyle[aas2pp4,psfig]{article}

\begin{document}

\title{\large\bf The Morphologically Divided Redshift Distribution of Faint
Galaxies}

\author{Myungshin Im \altaffilmark{1,6}, Richard E. Griffiths \altaffilmark{2},
 Avi Naim \altaffilmark{2}, Kavan U. Ratnatunga \altaffilmark{2},
 Nathan Roche \altaffilmark{3}, Richard F. Green \altaffilmark{4}, 
 \& Vicki L. Sarajedini \altaffilmark{5}}

\altaffiltext{1}{
 Space Telescope Science Institute, 
 Baltimore, MD 21218}

\altaffiltext{2}{
Dept. of Physics, Carnegie Mellon University, 
Pittsburgh, PA 15206}

\altaffiltext{3}{
 Dept. of Physics \& Astronomy, University of Cardiff, 
 P.O. Box 918, Cardiff CF2 3YB, Wales}

\altaffiltext{4}{
 NOAO, Tucson, AZ 85726-6732
}

\altaffiltext{5}{
 UCO/Lick Observatory, University of California, Santa Cruz, CA 95064}

\altaffiltext{6}{
 Current address, UCO/Lick Observatory, University of California, Santa Cruz, CA 95064
}


\begin{abstract}
    We have constructed a morphologically divided redshift distribution
 of faint field galaxies using a statistically unbiased sample of 196 galaxies 
 brighter than $I = 21.5$ for which detailed morphological information 
 (from the Hubble Space Telescope) as well as ground-based spectroscopic 
 redshifts are available.  Galaxies are classified into 3 rough morphological
 types according to their visual appearance (E/S0s, Spirals, Sdm/dE/Irr/Pec's),
 and redshift distributions are constructed for each type.  The most
 striking feature is the abundance of low to moderate redshift
 Sdm/dE/Irr/Pec's at $I < 19.5$.  This confirms that the faint
 end slope of the luminosity function (LF) is steep ($\alpha < -1.4$)
 for these objects.  We also find that Sdm/dE/Irr/Pec's are fairly
 abundant at moderate redshifts, and this can be explained by strong
 luminosity evolution.  However, the normalization factor (or the number
 density) of the LF of Sdm/dE/Irr/Pec's is not much 
 higher than that of the local LF of Sdm/dE/Irr/Pec's.  Furthermore, as we
 go to fainter magnitudes, the abundance of moderate to high redshift
 Irr/Pec's increases considerably. This cannot be explained by strong
 luminosity evolution of the dwarf galaxy populations alone: these
 Irr/Pec's are probably the progenitors of present day ellipticals
 and spiral galaxies which are undergoing rapid star formation or
 merging with their neighbors.  On the other hand, the redshift
 distributions of E/S0s and spirals are fairly consistent those expected 
from passive luminosity evolution, and are only in slight disagreement
 with the non-evolving model.

\end{abstract}


{\it Submitted to ApJ, Aug. 4th, 1997, Accepted July 27th, 1998}

\section{Introduction}

   The images of faint field galaxies taken with the Wide Field and
 Planetary Camera (WFPC2) on the Hubble Space Telescope (HST) have
 provided invaluable morphological information on these objects.  Using
 HST images with exposure times of about a few hours, reliable
 classification into basic morphological categories is possible down to
 a magnitude limit of $I \lesssim 22 $.  The Hubble Deep Field (HDF)
 observation, which is the deepest HST image so far, pushes this limit a
 few magnitudes fainter (Williams et al. 1996; Abraham et al. 1996; Naim
 et al. 1997).  Using these morphological classifications, it has been
 established with data from the HST Medium Deep Survey (MDS) and the HDF
 that the number counts at faint magnitudes are dominated by galaxies of
 irregular or peculiar appearance with small sizes (Im et al. 1995a,
 1995b; Griffiths et al. 1994a, 1994b; Casertano et al. 1995; Driver et
 al. 1995; Glazebrook et al. 1995).  Also, there is evidence that E/S0s
 and spiral galaxies have undergone passive Luminosity Evolution (LE) or
 have not evolved much since $z = 1$ (Im et al. 1996; Schade et
 al. 1997; Pahre et al. 1996; Bender et al. 1996).

Despite these important findings, it has not yet been shown decisively
as to what these faint irregular galaxies really are.  The number counts
and size distributions can be fitted well by assuming a model with a
steep faint end slope ($\alpha < -1.4$) for the LF of Sdm/Irr galaxies,
 and they could thus be irregular galaxies
with intrinsically low luminosity (Im et al. 1995a; Driver et al. 1995;
Glazebrook et al 1995).  But a considerable fraction (about 40 \%) of
faint Irr/Pec's show signs which can be interpreted as evidence for
interaction, suggesting that they could be merging galaxies at moderate
redshift (Driver et al. 1995).  Although the small sizes of these faint
galaxies suggest that they may not be high redshift $L_{*}$ galaxies
undergoing starburst activity (Im et al. 1995a; Roche et al. 1996), it
is not clear whether they are starbursting dwarf galaxies, or passively
evolving/starbursting sub $L_{*}$ spirals or E/S0s at $z \lesssim1$.
    
Fortunately, more spectroscopic redshifts (hereafter $z_{spec}$) are
becoming available for galaxies observed by the HST, thus providing a
fair sample of faint galaxies with morphological information as well as
$z_{spec}$.  As described in the next section, we have obtained about
120 redshifts for MDS galaxies brighter than $I = 21$.  Ground-based
follow-up spectroscopic observations have also been made for the HDF and
for other MDS samples (Cohen et al. 1996a, 1996b; Phillips et al. 1997;
Lowenthal et al. 1997; Forbes et al. 1996; Koo et al. 1996).  Also,
other groups have obtained HST WFPC2 images of galaxies in their
redshift surveys (Schade et al. 1995; Cowie et al. 1996), and these HST
data are now available for archival study .  The total number of faint
galaxies with spectroscopic redshifts and HST morphology now approaches
about 500, and the time is therefore ripe to construct the redshift
distributions for the morphologically divided faint galaxy samples.

\section{Data}

Our HST data include the HST MDS (Griffiths et al. 1994a), the strip
survey of Groth et al. (1994), the WFPC2 observations of three different
CFRS fields (Schade et al. 1995), the WFPC2 observations of the Hawaii
Deep field (Cowie et al. 1996) and the HDF itself.  The detection limit
for the fields with medium levels of exposure covers the range $I\simeq
24 \sim 25$, while the detection limit for the HDF goes as deep as $I
\simeq 28$.  For each object detected, the observed image is fitted with
simple model profiles (point source, $r^{1/4}$ profile and exponential
profile) using a 2-dimensional maximum likelihood technique (Ratnatunga
et al 1998a).  The resulting $I$-magnitude used here is a model-fit total
magnitude in the HST flight system using the F814W filter, and this is almost
equal to the conventional Johnson $I$ magnitude (Holtzman et al. 1995).

The morphologically classified galaxies are matched with the
spectroscopic redshift samples from Lilly et al. (1995a), Koo et
al. (1996), Forbes et al. (1996), Cohen et al. (1996b), Cowie et
al. (1996) and our own spectroscopic observations of MDS galaxies.  Our
own redshifts were obtained at the KPNO 4 meter telescope using the
technique of multi-object spectroscopy with the Cryocam.  Galaxies were
chosen primarily for their brightness.  Our limiting magnitude for
successfully measured redshifts was I$\simeq$21.0.  These spectra cover
the wavelength range from 4000$\AA$ to 9000$\AA$ with a resolution of
12$\AA$, and redshifts are based on emission or absorption
features. Adding these data to a collation of published results, a total
of 464 galaxies with spectroscopic redshifts are found with HST images
for morphological classification.  The authors (MI, AN, and NR) have
classified these galaxies according to their morphological appearance,
dividing them into three broad classes, i.e., E/S0s, Spirals, and
Sdm/dE/Irr/Pec galaxies.  The agreement level for this broad
morphological classification is about $90 \%$ at $I < 21.5$.  For the
purposes of this classification we have also used their luminosity
profiles as supplemental information to distinguish the dE population
from the normal E/S0 population (see Im et al. 1995b).  These dE's
should be distinguished from galaxies which are unclassifiable as a
result of their small sizes and faint magnitudes.  We treat
Sdm/dE/Irr/Pec galaxies as one galaxy type since their LFs have a steep
faint end slope, although we also find it plausible to treat Sdm/dE's
and Irr/Pec's as different galaxy types in our analysis (see section 4).
Coordinates, magnitudes, redshifts, and morphological classifications of
these galaxies will appear in a separate publication (Ratnatunga et
al. 1998b).

\section{Sample selection}

Our total sample of 464 galaxies is thus made up of heterogeneous
subsamples.  In order to construct a redshift distribution and to
compare it with the predictions of galaxy evolution models, it is
important to understand the completeness of each subsample.  We need to
make a correction for the fact that spectra were not taken for all the
galaxies in some fields. Consequently, the completeness of redshift
measurements varied from field to field, and thus a correction should be
made for this.  We define a quantity called ``redshift detection rate''
for this correction. This quantity is defined as the number of galaxies
with actual redshift measurements versus the number of galaxies with
photometric information for a given magnitude interval.  This quantity
must be distinguished with ``sampling rate'' which is defined as the
number of galaxies for which spectra are obtained over the number of
galaxies with photometric information.

  We divide our sample into two magnitude 
 bins $17.5 < I < 19.5$ (with the MDS sample) and $19.5 < I < 21.5$ 
 (without MDS sample) in order to eliminate biases which could affect
 the analysis. We find that there are 196 galaxies at $17.5 < I < 21.5$.
 LeFevre et al. (1995) discuss the completeness and the sampling 
 rate of the CFRS galaxies: the CFRS sample is about 90 \% 
 complete down to $I \simeq 21.5$ (or $I_{AB}=22$), and the sampling rate 
 for redshift measurements is about 22 \%.  They also discuss 
 potential biases resulting from spectral ranges, magnitudes,
 and surface brightness,  and find that their sample is not seriously
 biased by these quantities (Hammer et al. 1995).
 For the Hawaii Deep Field sample, Cowie et al. (1996) note that 
 their sample is about 93 \% complete down to the magnitude limit of
 $I=21.5$. Their sampling rate is 100 \% and thus the effect of color, 
 surface brightness or morphological bias is expected to be negligible. 
 Redshift measurements of HDF galaxies are nearly complete to I=21.5
 thanks to the collective efforts of several follow-up ground-based
 spectroscopic programs.
 Redshifts of galaxies in the HDF flanking fields have not been measured
 nearly as completely as those in the deep field. The sampling rate is
 about 50 \% down to I=22 for galaxies in the flanking fields. 
 For the MDS sample, we find that the completeness level varies depending on
 the observing runs. However, we note that redshift measurements  
 of galaxies in the MDS sample are nearly complete down to $I < 20$. 
 Therefore, we believe that our heterogeneous galaxy sample is free of
 bias at $I < 20$ within the MDS sample and to $I < 21.5$ without the
 MDS sample.  To show that our sample is not biased in terms of colors
 or sizes, we also present size-magnitude and color-magnitude diagrams.
 Figure 1 shows the size-magnitude diagram for 6 different subsamples
 (the results of three MDS spectroscopic follow-up runs separated by the
 year when the redshifts were taken; the Westphal-Groth strip; the
 Hawaii field; and the HDF follow-ups for the flanking fields).  We also
 present the color-magnitude diagrams for samples where more than one
 color is available (Figure 2).  If the sizes and colors of galaxies within
 a given magnitude range have a similar distribution to that of all the galaxies
 within the same magnitude range in that field (or fields), then those
 galaxies with redshifts are considered to have been randomly selected.
 Figures 1 and 2 show that there is no serious bias in terms of colors
 and sizes at the adopted magnitude limts.

\begin{figure}[tbh]
\centerline{\psfig{figure=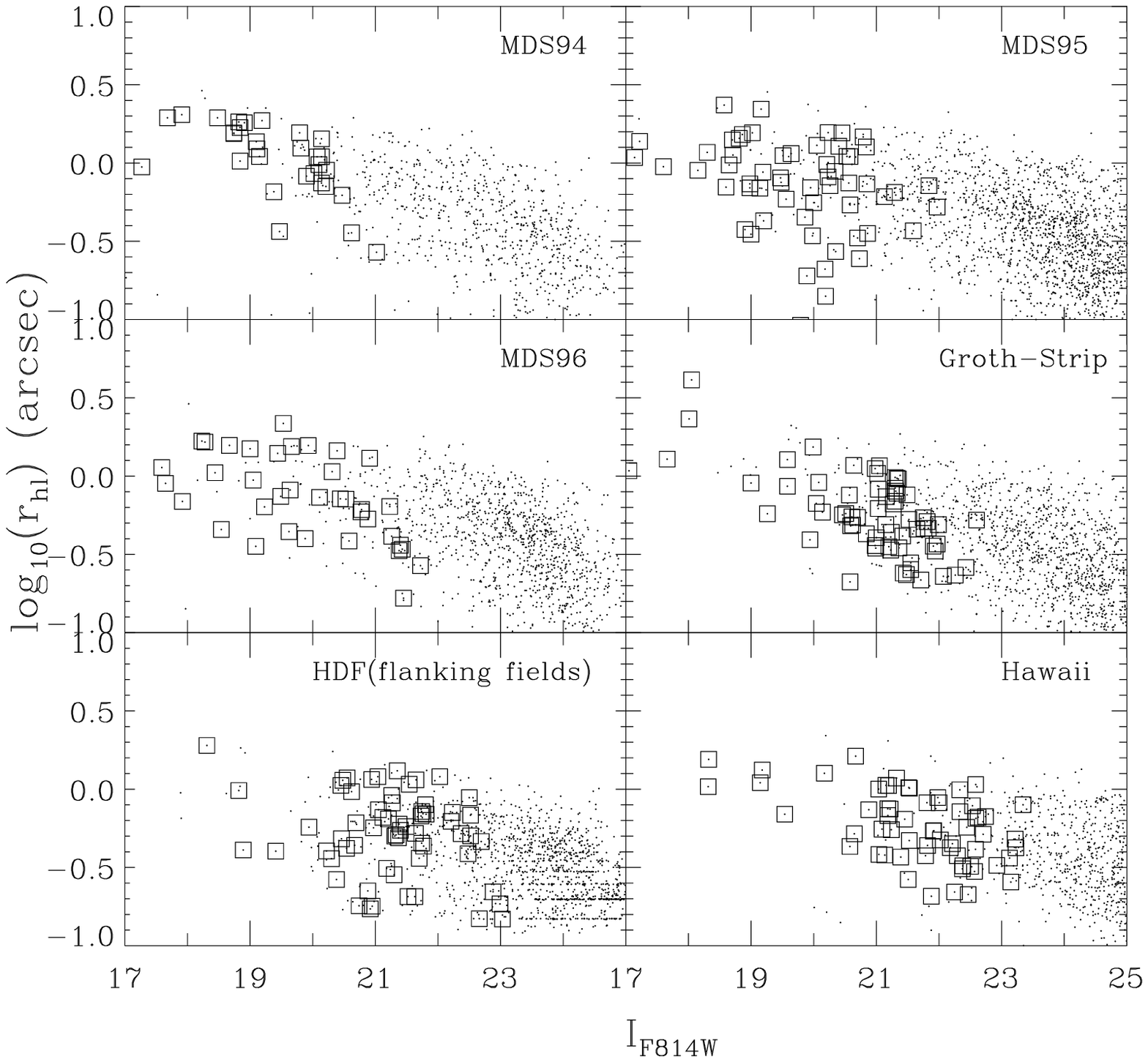,width=4.in}}
\footnotesize{
 Fig.1: The size-magnitude relation of various redshift samples (squares) 
       superposed on the relation for the total sample.}
\end{figure}

\begin{figure}[tb]
\centerline{\psfig{figure=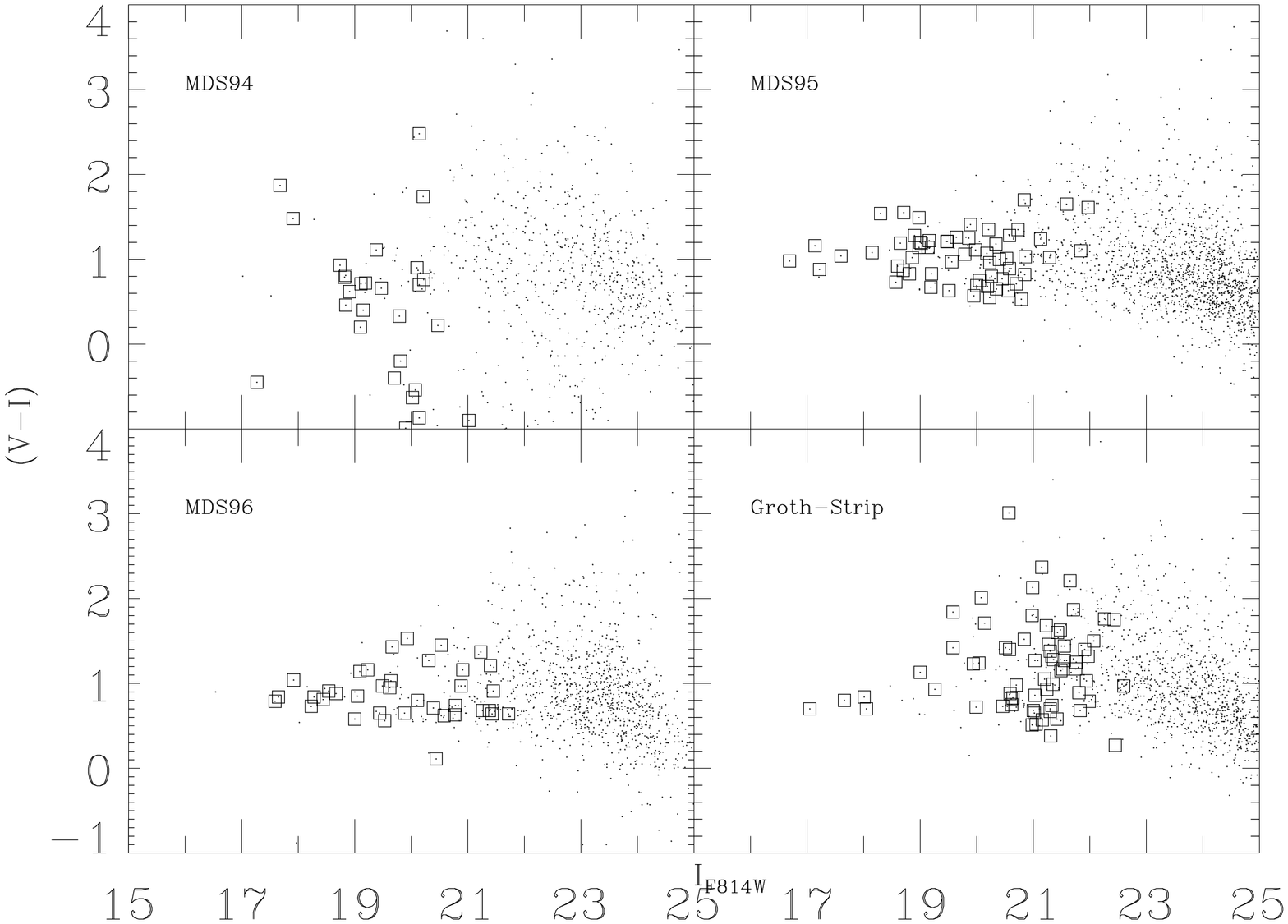,width=4.in,angle=90}}
\footnotesize{
 Fig.2: The color-magnitude relation of various redshift samples (squares) 
       superposed on the relation for the total sample.}
\end{figure}

 As a quantitative measure of the possible bias in the redshift sample,
 we have also studied the $V/V_{max}$ statistic for each galaxy type
 (Tables 1 - 3).  Originally proposed by Schmidt (1968), this quantity
 has been applied to various redshift samples to check if there is a
 bias in the sample or to find any evidence for cosmological evolution
 (e.g., see DellaCeca et al. 1992; Lilly et al. 1995b).  The basic idea
 is the following: first, the enclosed volume (V) is calculated.  The
 volume V is defined as the volume between the point where each galaxy
 is located and the closest point from the observer where a galaxy with
 the same absolute magnitude could possibly be found within the
 observational selection window (magnitude limit and redshift limit).
 Then, the volume V is compared with the maximum volume ($V_{max}$)
 where the galaxy could be found within the observational selection
 window.  If galaxies in the sample are randomly distributed in volume
 space, the quantity $V/V_{max}$s will have values randomly scattered
 between 0 to 1.  Consequently, the mean value of $V/V_{max}$s for the
 randomly selected sample will be 0.5, and any deviation from it would
 suggest some kind of bias in the sample which could originate from
 either an evolutionary or observational selection effect.
 Mathematically, $V/V_{max}$ is calculated using the following
 equations.
   
\begin{equation}
 V=\int^{z}_{max(z1,z_{m1})} (dV/dz)\,dz
\end{equation}

\begin{equation}
 V_{max}=\int^{min(z2,z_{m2})}_{max(z_{L},z_{m1})} (dV/dz)\,dz
\end{equation}

where z1 and z2 are the lower or upper limits of the redshift interval,
m1 and m2 are the lower or upper limits of the magnitude interval,
$z_{m1}$ and $z_{m2}$ are reshifts where the galaxy would be located if
it has apparent magnitudes m1 and m2 respectively, and (dV/dz) is the
volume element per unit redshift interval.

To obtain $z_{m1}$ or $z_{m2}$, we need to estimate the absolute
magnitude of each object and thus the (E+K) correction for each galaxy
type needs to be understood.  However, (E+K) corrections are not yet
well known.  Rather than using uncertain (E+K) corrections, we have used
K-corrections to estimate the absolute magnitude of each galaxy.  This
could lead to an underestimate of $V_{max}$ if there were luminosity
evolution (i.e, the brightening of galaxies as a function of redshift).
Hence, we would expect to get $<V/V_{max}>~>~0.5$ for a statistically
unbiased sample of passively evolving galaxies.  On the other hand, if
our sample is biased against detection of high redshift galaxies, we
expect to get $<V/V_{max}>~<~0.5$.

Tables 1, 2, and 3 show the values of $<V/V_{max}>$ for different types
of galaxies in different datasets and apparent magnitude ranges.  The
first number in parentheses is the actual number of galaxies with
redshifts and the second number has been corrected for sampling (see the
explanation below). All errors are 1-$\sigma$, ignoring any effects of
field to field fluctuations and small scale clustering. Thus, the real
errors are expected to be somewhat larger than those quoted.  Also note
that we estimated $<V/V_{max}>$ for a redshift interval of $0 < z < 1$.
Some caution must be taken in the interpretation of the $<V/V_{max}>$
values. Several of the surveys used here are slightly incomplete even at
this redshift and magnitude interval, and therefore $<V/V_{max}>$ could
be somewhat underestimated for some galaxy types.  For example, Hammer
et al.  (1997) note that it is very likely that redshifts of faint early
type galaxies at $z > 0.8$ are unidentified due to instrumental reasons.
Also, note that the $<V/V_{max}>$ values could fluctuate significantly
depending on the choice of the redshift interval, when galaxies are
distributed nonuniformly or spikily in redshift space like the E/S0s in
Fig.4.  Several caveats in the interpretation of $<V/V_{max}>$ values
are discussed in Im \& Casertano (1998).

The $<V/V_{max}>$ values for each galaxy type agree well with the
expected value of 0.5 - 0.6 within the errors, a result which is
consistent with luminosity evolution or no evolution.  Within the
fainter magnitude range ($19.5 < I < 21.5$), the $<V/V_{max}>$ values
appear to be greater than 0.5 for all galaxy types.  When galaxies
evolve passively without significant changes in their number density, we
expect $<V/V_{max}> \simeq 0.55$ if they are analysed assuming only
k-corrections (see, for example Im et al. 1996). The $<V/V_{max}>$ values of
all types of galaxies agree well with this expectation.  This would
tend to support the model of pure luminosity evolution of galaxies at moderate to high
redshift without strong number evolution, but this cannot be taken too
seriously since the errors are not sufficiently small.  Implications for
galaxy evolution based upon the $<V/V_{max}>$ values are discussed
in more detail in the next section.


In the next step, we estimated the redshift detection rate as a function
of apparent $I$ magnitude.  We defined the redshift detection rate to be
the number of galaxies with redshift divided by the total number of
galaxies in a given magnitude bin.  Table 4 shows the redshift detection
rate for the 7 subsamples as a function of $I$-band apparent magnitude.
The redshift detection rate may vary as a function of magnitude even in
the same subsample.  To construct the redshift distribution, we used the
inverse of the redshift detection rate to weight the number of galaxies
with redshifts in a given magnitude bin.

\section{Type-dependent redshift distribution}

The redshift distribution of E/S0s, Spirals and Sdm/dE/Irr/Pec's are
plotted as histograms in Fig.3 ($17.5 < I < 19.5$) and Fig.4 ($19.5 < I
< 21.5$).  Thin and thick lines show the distributions before and after
the application of the redshift detection rate correction.  Errors based
on poissonian statistics are shown for the thick lines.  Along with the
data, we have plotted the predicted redshift distribution using i) the
no evolution (NE) model (dashed line) and ii) the passive LE model
(solid line or dotted line).  The parameters for the LF of each type of
galaxy are listed in Table 5.  Especially for the Sdm/dE/Irr/Pec's, two
LE models are used according to the faint end slope of the LF, one for
$\alpha=-1.87$ (solid line) and one for $\alpha=-1.5$ (dotted line).  At
$17.5 < I < 19.5$, we used all the subsamples in Table 2, since they all
have magnitude limits fainter than $I=19.5$.  The most remarkable
feature in Fig. 3 is the abundance of Sdm/dE/Irr/Pec galaxies at low
redshift.  The redshift distribution for these galaxies peaks at $z
\lesssim 0.1$, and it is very difficult to obtain this kind of redshift
distribution without adopting a LF with a steep faint end slope,
confirming previous suspicions (Marzke et al. 1994; Gronwall \& Koo
1995; Im et al. 1995b).  In particular, we find that the number of these
Sdm/dE/Irr/Pec's is consistent with the prediction from the LF of Marzke
et al. (1994) within a factor of a few.  Further, the redshift
distribution of Sdm/dE/Irr/Pec's can be fit by assuming strong LE.  We
have used the LE model described by Driver et al. (1996).  For the
luminosity evolution parameter, we used $\beta = 0.7$.  The existence of
the $z \simeq 0.3 \sim 0.5$ Sdm/dE/Irr/Pec's is easy to understand in
the context of strong LE.  A similar conclusion has been reached from a
5-color survey of faint galaxies (Liu et al. 1997).  In contrast, the
redshift distributions of E/S0s and spirals are consistent with the
predictions of the passive LE model as well as those of the NE model.

\begin{figure}[pht]
\centerline{\psfig{figure=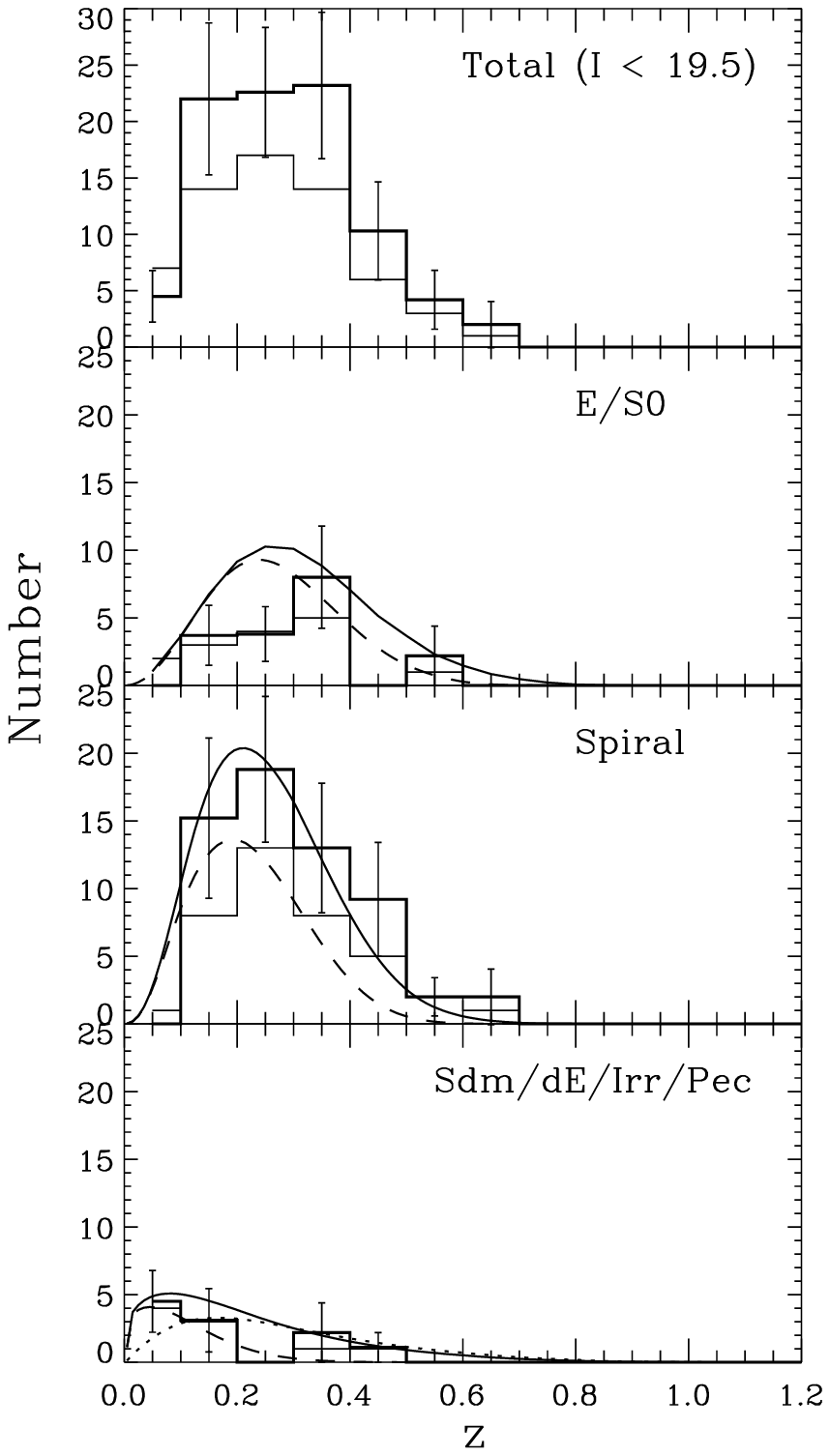}}
\footnotesize{
 Fig.3: The morphologically divided redshift distribution of galaxies 
      at $I < 19.5$. The solid histogram shows the distribution
      after the redshift detection rate correction is applied. The predicted 
      distribution from passive LE models are represented by the 
      solid line and the dotted line (model II for Sdm/dE/Irr/Pec),  
      and  the dashed line is for the NE model (see section 4). 
      The total number of
      galaxies used for this graph is 62, and the number of fields 
      is 54.}
\end{figure}

\begin{figure}[pht]
\centerline{\psfig{figure=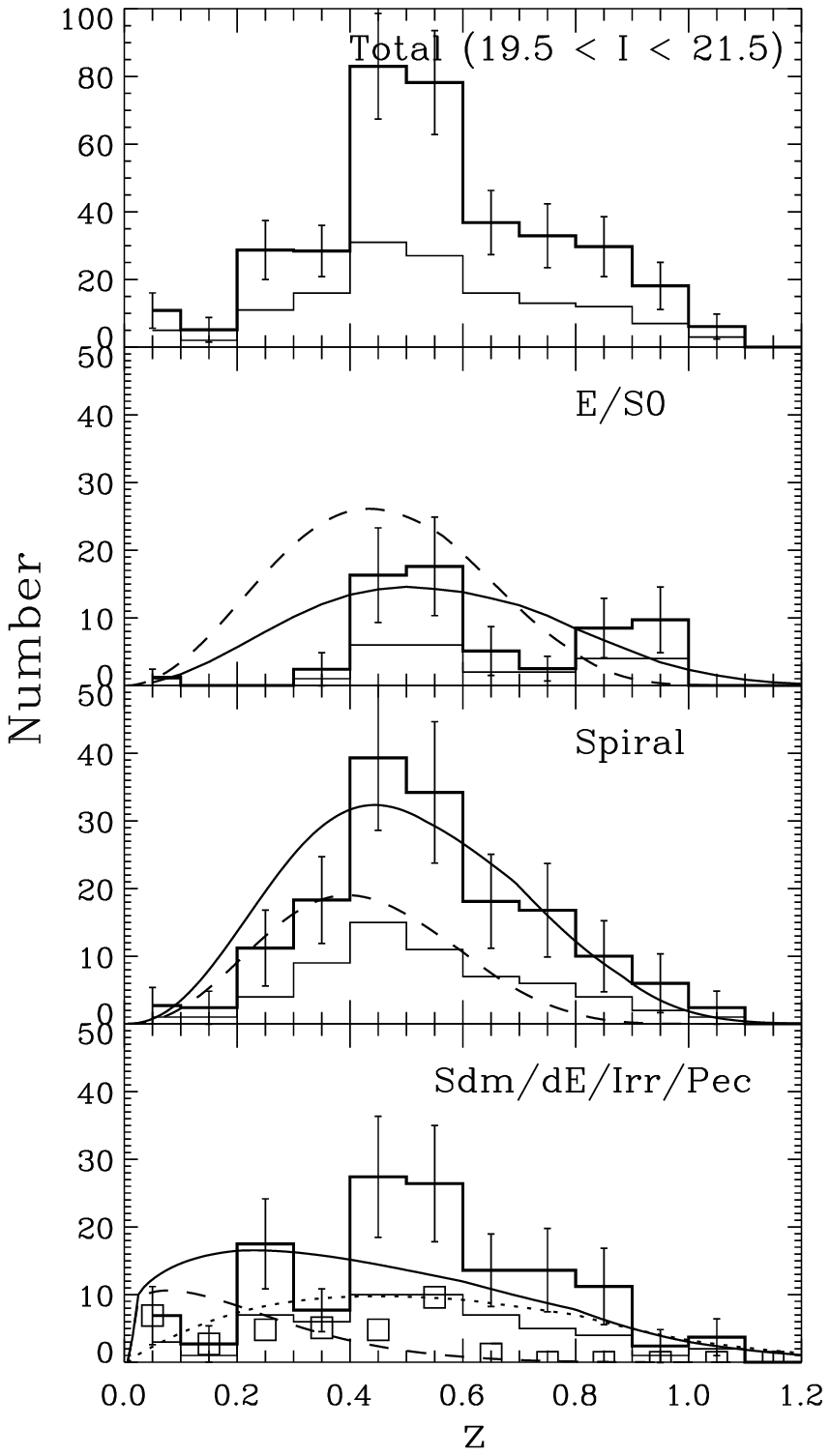}}
\footnotesize{
 Fig.4: The morphologically divided redshift distribution of galaxies 
      at $I < 21.5$.  The meaning of the lines is the same as Fig.3 
      except that the squares show the redshift distribution of
      Sdm/dE only with the redshift detection rate correction. 
      The total number of
      galaxies used for this graph is 143, and the number of fields 
      is 24. }
\end{figure}

At $19.5 < I < 21.5$, a different picture emerges for the
Sdm/dE/Irr/Pec galaxies. The MDS sample is excluded from the analysis in
this magnitude interval since the magnitude limit of the MDS sample is
$I \simeq 20$. 
No significant difference is found between the E/S0, spiral, and
Sdm/dE/Irr/Pec redshift distributions, contrary to the prediction of
dwarf-rich NE models where we would expect to find the peak of the
redshift distribution to be at $z \simeq 0.1$ (dashed line).  The LE
model also fails to match the observed redshift distribution (solid
line).  When the faint end slope of the LF is reduced to $\alpha = -1.5$
(dotted line, model II), the overall shape of the redshift distribution
can be matched better except for the normalization.  However, if we
force the normalization to fit, we predict too many Sdm/dE/Irr/Pec's at
the brighter magnitudes.

These results imply that: i) the Sdm/dE/Irr/Pec population cannot be
described by a simple LE or NE model with a steep faint end slope for
the LF, and that their evolution was much more complex; or that ii) the
majority of apparently Sdm/dE/Irr/Pec galaxies are star-forming or
interacting normal spirals or ellipticals, while some of the galaxies
classified as Sdm/dE/Irr/Pec are actually accounted for by a simple
dwarf rich LE model which occupy the low redshift domain of the
distribution.  To investigate hypothesis ii), we divided the
Sdm/dE/Irr/Pec galaxies into two different populations, one which
contains smooth objects (Sdm/dE) and the other which are not smooth
(Irr/Pec).  The redshift distribution of the smooth population is marked
with rectangles in Fig.4.  The smooth population accounts for the majority
of the low redshift sample, while the non-smooth population is
responsible for all the high redshift Sdm/dE/Irr/Pec objects ($z >
0.6$). Thus, most of the moderate to high redshift Sdm/dE/Irr/Pec's must
be objects which are experiencing violent activity such as starbursts or
merging.  Verification of hypothesis i) is not a trivial task.  As a
simple prescription, we model the number evolution of Sdm/dE/Irr/Pecs as
being proportional to $(1+z)^m$.  Since the predicted numbers of
Sdm/dE/Irr/Pec's are low by a factor of 2 - 3 at $z > 0.4$, it will be
sufficient to adopt number evolution rising as $(1+z)^{2 \sim 3}$ in
order to fit the observed distribution.

The $<V/V_{max}>$ value could, in principle, be a sensitive indicator of number
evolution. However, since errors are not sufficiently small, the
$<V/V_{max}>$ values are consistent with both $\sim 0.55$ (strong
luminosity evolution only) and $\sim 0.6$ (strong luminosity evolution
plus number density evolution). Based on this test, it is difficult to
judge which hypothesis is right.

The question still
remains as to the nature of these moderate to high redshift Irr/Pec's:
are they $L_{*}$ - $sub\,L_{*}$ spirals that are forming stars more
actively than the present day spirals, or are they the present-day dwarf
galaxies which were in a starburst stage (Babul \& Ferguson 1996) and
disappeared later?  To arrive at a full answer, other observables will be
helpful, such as colors, sizes and velocity dispersions.  A detailed
analysis of colors, sizes and redshifts of these galaxies has been
conducted by Roche et al. (1998).  The  indications from this 
analysis are such
that the colors of the Irr/Pec's show a wide dispersion on the
color-redshift diagram indicating that some of these galaxies are more
consistent with being spirals or E/S0s.  Furthermore, the
size-luminosity relation of Irr/Pec's at different redshifts indicates
that simple LE is not enough to explain their compactness, and that
strong LE or size evolution is necessary for some Irr/Pec's.  There is
also evidence for the existence of starbursting $L_{*}$ - $sub \, L_{*}$
galaxies at $z \gtrsim 0.3$ from a 5-color photometric survey of faint
galaxies (Liu et al. 1997).  These pieces of evidence appear to favor
hypothesis (ii), so that strong number evolution of the Irr/Pec
population is not necessary.

Finally, we note that the redshift distributions of E/S0s and spiral
galaxies at $19.5 < I < 21.5$ are consistent with the prediction of the
passive LE model and the NE model. The NE models appear to fail to
predict the right abundance of E/S0s and spirals at $z < 0.7$, but the
difference may not be significant due to the uncertainty in the
normalization of their LFs.  A notable feature is the spiky nature of
the distribution of E/S0s. This is expected since E/S0s are more
clustered than other types of galaxies (e.g, Neuschaefer et al. 1997).

The $<V/V_{max}>$ values of E/S0s and spirals are consistent with values
greater than 0.5, supporting the idea that these galaxies evolved
passively.  In particular, our $<V/V_{max}>$ value for E/S0 galaxies is
significantly greater than 0.4 at $19.5 < I < 21.5$. This contradicts
the result of Kauffmann et al. (1996) in which they reported a strong
number evolution of early type galaxies at $z < 1$ ($\sim
(1+z)^{-1.5}$). Their claim is based on the measurement of $<V/V_{max}>
\simeq 0.4$ for their color-selected early type galaxies.  Our result
from the morphologically selected E/S0 sample does not support such
strong number evolution, confirming instead the earlier result from a
much larger sample of morphologically selected E/S0s with photometric
redshfits (Im et al. 1996) where we reported $<V/V_{max}> \simeq
0.55-0.58$.  This indicates that the number density of E/S0s has not
changed significantly since z=1 (also see, Totani \& Yoshii 1998;
 Im \& Casertano 1998). 


\section{Conclusions}

We have constructed morphologically divided redshift distributions of
$\sim 200$ galaxies in two magnitude intervals, $I < 19$, and  $19.5 < I <
21.5$.  Redshifts of these galaxies are taken from
(largely published) spectroscopic observations and the morphological
classification has been done using HST data.  The observed redshift
distribution of Sdm/dE/Irr/Pec's at $I < 19.5$ indicates that 
a very high normalization for the LF of Sdm/dE/Irr/Pec's is {\it unnecessary},
but the LF does need to have a steep faint end slope, 
confirming the findings from catalogs of nearby galaxies (Marzke et al. 1995).
We also find that there was strong luminosity evolution for
Sdm/dE/Irr/Pec's, but that the strong LE of Sdm/dE/Irr/Pec's alone is
not enough to explain the moderate to high redshift Irr/Pec's at $I >
19.5$.  Many Irr/Pec's at moderate to high redshift must be either
starbursting spirals and E/S0s or disappearing dwarfs with a number
density evolution of $\sim (1+z)^2$.
The preliminary analysis of colors, sizes and redshifts of these
galaxies indicates that many Irr/Pec's are likely to be sub $L_{*}$
galaxies rather than starbursting dwarf galaxies.  In contrast with
this situation for Irr/Pec's, we find that the observed redshift
distributions of E/S0s and spirals are consistent with the various
evolutionary models, and do not require strong number density evolution at $z < 1$.

\acknowledgements

We thank an anonymous referee for useful comments and a careful review
of the paper, and Pete Stockman for supporting this work. 
 This paper is based on observations with the NASA/ESA
Hubble Space Telescope, obtained at the Space Telescope Science
Institute, which is operated by the Association of Universities for
Research in Astronomy, Inc., under NASA contract NAS5-26555.  The HST
Medium Deep Survey has been funded by STScI grants GO2684 {\it et seqq.}
Also, this work is partly supported by the STScI Director's 
Discretionary Research Fund.

\newpage

\begin{deluxetable}{c c c}
\scriptsize
\tablecolumns{3}
\tablecaption{$<V/V_{max}>$ for each galaxy type}
\tablewidth{0pc}
\tablehead{
 Galaxy  & \multicolumn{2}{c}{Magnitude Bin} \\ \cline{2-3} \\
  Type   & \colhead{$17.5 < I < 19.5$} & \colhead{$19.5 < I < 21.5$}} 
\startdata
 Total    & $0.54 \pm 0.04$ (55,85)  & $0.55 \pm 0.02$ (141,355) \nl
 E/S0     & $0.40 \pm 0.08$ (13,17)  & $0.60 \pm 0.05$ (26,63)   \nl
 Spirals  & $0.60 \pm 0.05$ (35,58)  & $0.53 \pm 0.04$ (61,162)  \nl
Sdm/dE/Im & $0.59 \pm 0.20$ (2,3)    & $0.57 \pm 0.07$ (17,35)   \nl
 Pec/Irr  & $0.38 \pm 0.13$ (5,7)    & $0.55 \pm 0.05$ (37,94)   \nl
\enddata
\end{deluxetable}

\begin{deluxetable}{l c c c c c c c}
\tiny
\tablecolumns{8}
\tablecaption{$<V/V_{max}>$ for each survey ($17.5 < I < 19.5$)}
\tablewidth{0pc}
\tablehead{
 Galaxy  & \multicolumn{7}{c}{Survey Name} \\ \cline{2-8} \\
  Type   & \colhead{MDS94} & \colhead{MDS95} & \colhead{MDS96} 
         & \colhead{Lilly+Groth Strip} 
         & \colhead{HDF} & \colhead{HDF(flanking)} 
         &  \colhead{Hawaii}}  
\startdata
 Total    & $0.46 \pm 0.08$ (11,18) & $0.59 \pm 0.07$ (15,21) 
          & $0.61 \pm 0.09$ (10,18) & $0.46 \pm 0.09$ (9,15) 
          &  $ 0.50 \pm 0.14$ (4,4) 
          & $0.78 \pm 0.20$ (2,4)   & $0.44 \pm 0.14$ (4,4)    \nl
 E/S0     & $0.20 \pm 0.14$ (4,6)   & $0.38 \pm 0.16$ (3,3)
          &  \nodata                & $0.76 \pm 0.20$ (2,4)
          & $0.47 \pm 0.28$ (1,1) 
          &  \nodata                & $0.20 \pm 0.20$ (2,2)    \nl
 Spirals  & $0.60 \pm 0.11$ (7,12)  & $0.65 \pm 0.08$ (11,17)
          & $0.58 \pm 0.09$ (9,16)  & $0.30 \pm 0.16$ (3,5) 
          & $0.69 \pm 0.20$ (2,2)
          & $0.78 \pm 0.28$ (1,4)   & $0.67 \pm 0.20$ (2,2)    \nl
 Sdm/dE   &  \nodata                &  \nodata
	  & $0.83 \pm 0.28$ (1,2)   & $0.10 \pm 0.28$ (1,1)
          &  \nodata      
          &  \nodata                &  \nodata                 \nl
 Pec/Irr  &  \nodata                & $0.35 \pm 0.28$ (1,1)
	  &  \nodata                & $0.43 \pm 0.16$ (3,5) 
          & $0.16 \pm 0.28$ (1,1)
          &  \nodata                &  \nodata                 \nl
\enddata
\end{deluxetable}

\newpage

\begin{deluxetable}{c c c c c}
\scriptsize
\tablecolumns{5}
\tablecaption{$<V/V_{max}>$ for each survey ($19.5 < I < 21.5$)}
\tablewidth{0pc}
\tablehead{
 Galaxy  & \multicolumn{4}{c}{Survey Name} \\ \cline{2-5} \\
  Type   & \colhead{Lilly+Groth Strip} 
         & \colhead{HDF} & \colhead{HDF(flanking)} 
         &  \colhead{Hawaii}}  
\startdata
 Total    & $0.53 \pm 0.03$ (66,166) & $0.59 \pm 0.06$ (24,51) 
          & $0.56 \pm 0.05$ (33,115) & $0.58 \pm 0.07$ (18,23) \nl 
 E/S0     & $0.60 \pm 0.10$ (8,20)   & $0.69 \pm 0.10$ (9,19)
          & $0.54 \pm 0.11$ (6,21)   & $0.57 \pm 0.16$ (3,4)   \nl
 Spirals  & $0.45 \pm 0.05$ (28,72)  & $0.62 \pm 0.09$ (10,24)
          & $0.58 \pm 0.07$ (17,60)  & $0.58 \pm 0.11$ (6,7)   \nl
 Sdm/dE   & $0.63 \pm 0.10$ (8.20)   & $0.04 \pm 0.20$ (2,2)
	  & $0.50 \pm 0.20$ (2,7)    & $0.63 \pm 0.12$ (5,6)   \nl
 Pec/Irr  & $0.56 \pm 0.06$ (22,55)  & $0.39 \pm 0.16$ (3,6)
	  & $0.56 \pm 0.10$ (8,28)   & $0.54 \pm 0.14$ (4,5)   \nl 
\enddata
\end{deluxetable}

\begin{deluxetable}{l c c c c c c c l}
\scriptsize
\tablecolumns{9}
\tablecaption{Redshift Detection Rates}
\tablewidth{0pc}
\tablehead{ \colhead{I mag} & \colhead{MDS94} & \colhead{MDS95} & 
\colhead{MDS96} & \colhead{CFRS(+ GROTH strip)} & \colhead{Koo et al.} & 
\colhead{HDF} & \colhead{HDF(flanking)} & \colhead{Hawaii} } 
\startdata
 17.5  & 1.00  & 1.00  & 0.68  &  1.00  &  \nodata &    1.00  &  0.00  & 1.00   \nl
 18.5  & 0.71  & 0.90  & 0.68  &  0.33  &   0.68   &  \nodata &  0.43  & 1.00   \nl
 19.5  & 0.47  & 0.49  & 0.56  &  0.26  &   0.00   &    1.00  &  0.25  & 1.00   \nl
 20.5  & 0.23  & 0.31  & 0.28  &  0.32  &   0.20   &    0.94  &  0.30  & 0.71   \nl
 21.5  & 0.00  & 0.04  & 0.12  &  0.39  &   0.12   &    0.80  &  0.28  & 0.71  \nl
 22.5  & 0.00  & 0.00  & 0.00  &  0.03  &   0.05   &    0.41  &  0.06  & 0.34   \nl
 23.5  & 0.00  & 0.00  & 0.00  &  0.00  &   0.00   &    0.05  &  0.00 & \nodata   \nl
\enddata
\end{deluxetable}

\begin{deluxetable}{l c c c c c}
\tiny
\tablecolumns{5}
\tablecaption{Model Parameters}
\tablewidth{0pc}
\tablehead{ \colhead{Galaxy type} & \colhead{$\alpha$} & 
\colhead{$M_{*}(I)\,+\,5\,log_{10}(h)$} & 
\colhead{$\phi_{*} (h^{3}~Mpc^{-3})$} & 
\colhead{Star formation history} } 
\startdata
 E/S0     & -0.92  & -21.5  & 0.005  &  1 Gyr burst, Salpeter IMF,\& $z_{for}=5$ (Bruzual \& Charlot 1996) \nl
 Spirals  & -0.92  & -20.9  & 0.015  &   $\mu=0.25$(exponential), Scalo IMF, \& $z_{for}=3$ (Bruzual \& Charlot 1996)  \nl
 Sdm/dE/Irr/Pec (I) & -1.87  & -19.6  & 0.004  & 
  Evolution model of Driver et al. (1996) \nl
 Sdm/dE/Irr/Pec (II) & -1.50  & -19.6  & 0.004  & 
  Evolution model of Driver et al. (1996) \nl
\enddata
\end{deluxetable}

\end{document}